\documentclass[12pt]{article}
\usepackage{newtxtext,newtxmath}
\usepackage{graphicx}
\usepackage[letterpaper,margin=1in]{geometry}
\renewenvironment{abstract}
	{\quotation}
	{\endquotation}
\date{}

\makeatletter
\renewcommand{\fnum@figure}{\textbf{Figure \thefigure}}
\renewcommand{\fnum@table}{\textbf{Table \thetable}}
\makeatother

\usepackage{scicite}
\usepackage{url}

\def\scititle{
Toward ultimate-efficiency frequency conversion in nonlinear optical microresonators
}
\title{\bfseries \boldmath \scititle}
\author{
Zhi-Yan Wang$^{1\dagger}$,
Xiao Wu$^{2\dagger}$,
Xiao Xiong$^{1}$,
Chen Yang$^{1}$,
Zhengzhong Hao$^{2}$,\\
Qi-Fan Yang$^{1,4}$,
Yaowen Hu$^{3}$,
Fang Bo$^{2\ast}$,
Qi-Tao Cao$^{1\ast}$,
Yun-Feng Xiao$^{1,4\ast}$\and
\small$^{1}$State Key Laboratory for Mesoscopic Physics, Frontiers Science Center for Nano-optoelectronics, \\ \small New Cornerstone Science Laboratory, School of Physics, Peking University, 100871 Beijing, China\and
\small$^{2}$MOE Key Laboratory of Weak-Light Nonlinear Photonics, \\ \small TEDA Institute of Applied Physics and School of Physics, Nankai University, Tianjin 300457, China\and 
\small$^3$John A. Paulson School of Engineering and Applied Sciences,\\ \small Harvard University, Cambridge, MA, USA\and
\small$^4$Collaborative Innovation Center of Extreme Optics, Shanxi University, Taiyuan 030006, China.\and
\small$^\ast$Fang Bo: bofang@nankai.edu.cn\and
\small$^\ast$Qi-Tao Cao: caoqt@pku.edu.cn\and
\small$^\ast$Yun-Feng Xiao: yfxiao@pku.edu.cn\and
\small$^\dagger$These authors contributed equally to this work.
}

\begin{document}

\maketitle

\begin{abstract} 
\bfseries \boldmath 
Integrated nonlinear photonics has emerged as a transformative platform, enabling nanoscale nonlinear optical processes with significant implications for sensing, computation, and metrology. 
Achieving efficient nonlinear frequency conversion in optical microresonators is paramount to fully unlocking this potential, yet the absolute conversion efficiency (ACE) of many processes, such as second-harmonic generation (SHG), remains fundamentally constrained by dissipative losses and intrinsic nonlinear effects in the device. 
In this work, we establish a unified theoretical framework for SHG in microresonators, identifying a decisive factor $M$ that predicts the upper limit of ACE under the nonlinear critical coupling (NCC) condition.
Using this framework, we fabricate integrated periodically poled lithium niobate microresonators and address the dispersive and dissipative suppression to approach the NCC condition.
We achieve a record-high experimental ACE of 61.3\% with milliwatt-level pump powers toward the ultimate efficiency, with the potential for even higher efficiency as the $M$ factor increases.
These results provide a versatile paradigm for high-efficiency nonlinear optical devices, offering new opportunities for advancements across classical and quantum photonic applications.
\end{abstract}

\noindent

\subsection*{Introduction} 

The observation of frequency-doubled signals in laser-illuminated quartz crystals marked a pivotal development in the advancement of nonlinear optics \cite{1961generation}, establishing a foundation for numerous photonics applications. Early demonstrations of nonlinear optical processes were constrained by low conversion efficiencies, a challenge fundamentally overcome with the advent of phase-matching theory. By ensuring the conservation of energy and momentum, phase matching enables theoretically complete conversion of light into harmonic frequencies over sufficiently long interaction lengths \cite{boyd2020nonlinear, frequencyconversion}. 
This advancement enables the creation of highly efficient harmonic laser sources spanning the visible to deep ultraviolet spectrum, driving progress in ultrafast spectroscopy \cite{attosecondphysics,uv,dualcomb,ultrafast}, quantum optics \cite{qtharmonicl,qtharmonicl2,qtharmonicl3,qtharmonicl4}, and precision metrology \cite{metrology,clock,metro}.

Further enhancing harmonic generation efficiency often involves embedding nonlinear media into optical microresonators, for their offering considerable advantages over their bulk counterparts \cite{2004ppln,2010selfpulsing,yupinghuang,wuxiao,AlN2018}. 
First, the small-scale cross-sections of microresonators generate higher optical intensities than free-space beams at equivalent power levels. 
Second, microresonators with high-quality ($Q$) factors and large free spectral ranges (FSRs) provide thousands of light recirculations, effectively increasing the interaction length within a compact footprint. 
These advantages greatly reduce the power requirements for many nonlinear optical processes. 
Consequently, microresonator-based harmonic generators constructed from high-nonlinearity materials \cite{jintianlin,lu2019efficient,single,fangkejie,reconfigurable,luphcr,zouclsfg} have become fundamental components in integrated photonics \cite{geatareview}, enabling complex communication and computational tasks through synergistic integration with other photonic devices.

In practical applications, the absolute power of harmonic lasers is critical, making absolute conversion efficiency (ACE)---defined as the ratio of converted power to input power---a key performance metric. However, achieving near-unity ACE in microresonators remains challenging for both second- and third-order nonlinear processes though the phase matching condition is taken into consideration \cite{elight,opoefficiency,single}. 
For example, in second harmonic generation (SHG), integrated microresonators based on periodically poled lithium niobate (PPLN) have achieved a relative conversion efficiency up to $5{,}000{,}000\%/\text{W}$ with microwatt input powers.
Nonetheless, as the input power increases, the accessible ACE is very limited, which is a common result for microcavity-enhanced systems, with the efficiency plateauing around $30\%$ \cite{ALN2500, PPLN250000, luxiyuan, 2022cascaded, linqiang}.  
The theoretical upper limit of the conversion efficiency is being pursued, yet it remains unclear what the comprehensive suppression mechanisms and critical conditions for achieving this limit entail.
In this article, we present a unified theoretical framework figuring out the suppression of ACE to realize the ultimate-efficiency energy conversion, where a decisive factor $M$ for ACE limit and the corresponding strict condition called the nonlinear critical coupling (NCC) condition is proposed.  
The theory is experimentally validated using a high-$Q$ PPLN-integrated microresonator, wherein we demonstrate an on-chip SHG efficiency of $61.3\%$ at milliwatt pump powers, representing a record high among microcavity-enhanced photonic platforms.

\subsection*{Results}

\subsection*{Fundamental limitations of ACE} 

\begin{figure}
\centering
\includegraphics{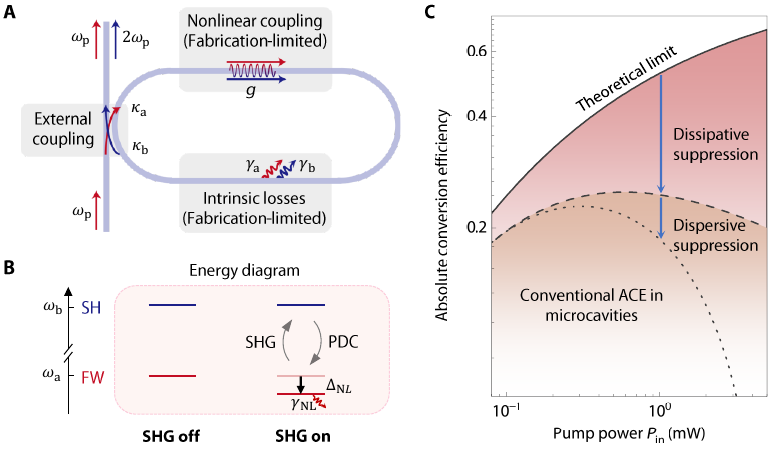}
\caption{\textbf{SHG in nonlinear microresonators.}
\textbf{(A)} Schematic illustration of the microresonator-based SHG process. The parameters are either limited by fabrication or tunable during design and experimentation.
\textbf{(B)} Energy-level diagram for modes involved when the SHG process in a microcavity is off and on. PDC: parametric down-conversion. 
\textbf{(C)} Schematic of the ACE versus pump power under different conditions, showing the dispersive and dissipative suppression.
Gray solid (dashed) curve denotes the ACE of the theory limit (the double-resonance condition).
Gray dotted curve depicts the measured ACE considering the nonlinear modulations within the cavity.
}
\label{fig1} 
\end{figure}

SHG within a nonlinear microresonator involves two optical modes: the fundamental wave (FW) mode $a$ and the second harmonic (SH) mode $b$ (Fig.~\ref{fig1}A).
The single-photon nonlinear coupling strength is denoted by $g$ and the intrinsic losses of these modes are characterized by $\gamma_\mathrm{a}$ and $\gamma_\mathrm{b}$, respectively. The phase-matching condition is assumed to be satisfied. 
Ideally, a larger $g$ and a smaller $\gamma$ are desirable, which are constrained by practical fabrication limitations.
Besides, the external coupling losses $\kappa_{\rm a}$ and $\kappa_{\rm b}$ affect the input and output efficiencies for the pump and harmonic light.
The SHG process creates a nonlinear coupling between the FW and SH modes, with a complex and intensity-dependent coupling strength.
This process can be quantitatively explored by solving the nonlinear coupled-mode equations (see Supplementary Materials), and the evolution of the FW field can be derived as
\begin{equation}
\frac{\mathrm{d}a}{\mathrm{d}t} = i \left(\Delta_\mathrm{a}+\Delta_\mathrm{NL}\right) a - \frac{\kappa_\mathrm{a}+\gamma_\mathrm{a}+\gamma_\mathrm{NL}}{2} a + \sqrt{\frac{\kappa_\mathrm{a} P_\mathrm{in}}{\hbar \omega_\mathrm{p}}},
\label{Eq:modFW} 
\end{equation}
where $a$ represents the field amplitude of the FW mode. 
The detunings are defined as $\Delta_\mathrm{a} = \omega_\mathrm{a} - \omega_\mathrm{p}$ and $\Delta_\mathrm{b} = \omega_\mathrm{b} - 2\omega_\mathrm{p}$, and $\omega_\mathrm{p}$ is the pump frequency.
There exist some other conventional nonlinear processes, such as the photothermal effect, Kerr nonlinearities, and the photorefractive effect, which are incorporated into $\omega_\mathrm{a}$ and $\omega_\mathrm{b}$.
Notably, the SHG and parasitic parametric down-conversion (PDC) processes \cite{linranfan} bring the FW mode an additional frequency shift
\begin{equation}
\Delta_\mathrm{NL} = -\frac{2 g^2 |a|^2 \Delta_\mathrm{b}}{\Delta_\mathrm{b}^2 + \left(\kappa_\mathrm{b}+\gamma_\mathrm{b}\right)^2 / 4},
\label{Eq:nfreqshift}
\end{equation}
and an extra loss
\begin{equation}
\gamma_\mathrm{NL} = \frac{2g^2 |a|^2 \left(\kappa_\mathrm{b}+\gamma_\mathrm{b}\right)}{\Delta_\mathrm{b}^2 + \left(\kappa_\mathrm{b}+\gamma_\mathrm{b}\right)^2 / 4},
\label{Eq:nloss}
\end{equation}
of which the effects are schematized in Fig. \ref{fig1}B.

Both the shift and loss are proportional to $|a|^2$, leaving strengthened dispersive and dissipative modulation for the FW mode with the increasing intracavity power.
Thus, although resonance relatively enhances the nonlinear process in comparison to the non-resonant system, these modulations respectively weaken the intracavity energy conversion and degrade the input efficiency of the pump, fundamentally suppressing the ACE from reaching the theoretical limit.
As qualitatively illustrated in Fig. \ref{fig1}(C), dispersive and dissipative suppressions contribute comparably and become more significant with increasing pump power, imposing a critical limitation on applications that require substantial pump power.
In order to reach the theoretical upper limit of ACE, the dispersive suppression is required to be overcome within the experiments as well as the dissipative suppression through engineering guidance. 

We further evaluate the ACE limit and the corresponding condition to overcome the dispersive-dissipative suppression.
The ACE in this SHG system is defined as the ratio of output SH power to the input pump power for the cavity, which is determined by the intrinsic parameters $g$ and $\gamma_{\rm{a(b)}}$, experimental parameters $\Delta_{\rm{a(b)}}$, and external coupling parameters $\kappa_{\rm{a(b)}}$.
Through further analytical derivation (see Supplementary Materials), the maximum ACE is expressed as \begin{equation}
    \mathrm{ACE_{\max}} = 1-\frac{1}{R(M)}-\frac{(R(M)-1)^2}{M},
    \label{ACElimit}
\end{equation}
where the dimensionless $M$ is a decisive factor, presenting
\begin{equation}
    M = \frac{8g^2 P_{\mathrm{in}}}{\hbar \omega_{\mathrm{p}} \gamma_\mathrm{a}^2 \gamma_\mathrm{b}}.
\end{equation}
$R(M)$ is a function of $M$ factor (see Supplementary Materials).
This maximum ACE is obtained only when double resonance is realized and the relationship for the external coupling rates for both modes satisfies $\kappa_{\rm a}/\gamma_{\rm a}=(\kappa_{\rm b}+\gamma_{\rm b})/2\gamma_{\rm b}=R(M)$ as well.
Notably, it indicates that the $M$ factor directly brings the upper limit of ACE and the required external coupling rates, while the $M$ factor is only determined by the intrinsic parameters and pump power.
Consistent and reproducible fabrication quality enables reliable prediction of the intrinsic parameters, thereby informing the experimental operation and microresonator design.

The condition for achieving optimal ACE within such a two-mode nonlinear coupling model is further analyzed.
The derivation reveals that this optimization condition corresponds to zero transmission for the FW mode, which, importantly, is only a necessary condition for reaching the limit (see Supplementary Materials).
A strict constraint on the external coupling rate for the SH mode is also required.
To summarize, the critical condition for maximum ACE writes
\begin{align}
&\Delta_\mathrm{a} + \Delta_\mathrm{NL} = 0, \Delta_\mathrm{b}=0 \\
&\kappa_\mathrm{a} = \gamma_\mathrm{a} + \gamma_\mathrm{NL}, \\
&(\kappa_{\rm a}+\gamma_{\rm{NL}})/\gamma_{\rm a}=\kappa_{\rm b}/\gamma_{\rm b},
\label{Eq:GCC}
\end{align}
which is called nonlinear critical coupling (NCC) condition. 
It is a stricter condition compared to the generalized critical coupling condition for linear coupling system \cite{yaowenhu2,yaowenhu1} and it is different from the condition only bringing coupling constraints for the pumped mode \cite{ol}.
More intuitively, this condition corresponds to the scenario where the light frequency matches the modes and the external coupling rates align with the intrinsic losses and nonlinear coupling strength. 
It requires that all the pump power is injected into the system for a double-resonant energy conversion process, and then a maximum extraction efficiency for SH light.
The simple form for the relationship of external coupling rates for modes in Eq. \ref{Eq:GCC} exhibits an intuitive process: two photons in the FW mode flow from the waveguide to the SH mode at the same effective rate and become one photon, and then this photon flows from SH mode to the waveguide with the same rate.

\begin{figure}
\centering
\includegraphics{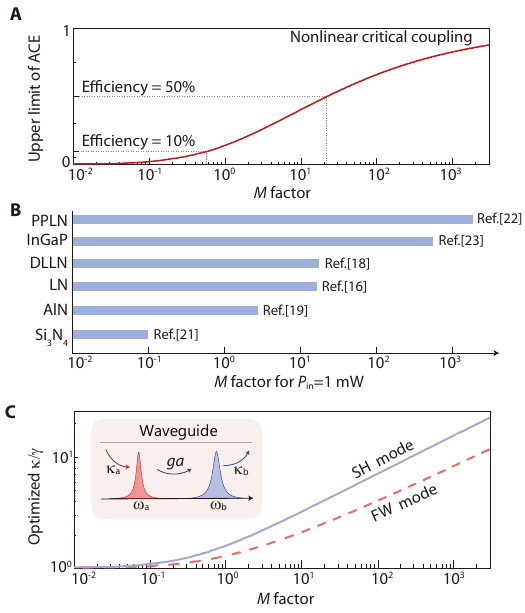}
\caption{\textbf{Upper limit of ACE and nonlinear critical coupling (NCC) condition in microresonators.}
\textbf{(A)} Upper limit of absolute conversion efficiency as a function of $M$. 
\textbf{(B)} The largest achievable $M$ for a pump power of $1$ mW is compared across various materials platforms. 
DLLN: double-layer lithium niobate.
\textbf{(C)} Ratio of external coupling rates to intrinsic losses optimizing for maximum ACE under the NCC condition as a function of $M$.
Inset: Schematic of energy flow within the double-resonant SHG process.}
\label{fig2} 
\end{figure}

Quantitatively, Fig.~\ref{fig2}A illustrates the relationship between the ACE limit and $M$ factor. 
To attain a substantial maximum ACE ($>10\%$), $M$ needs to approach unity. 
When near-unity efficiency is required, $M$ needs to exceed thousands.
We also provide a comprehensive list of demonstrated nonlinear microresonator platforms and their attainable $M$ in Fig. \ref{fig2}B. 
To date, PPLN-integrated microresonators lead the competition in achieving high $M$, demonstrating the greatest potential for attaining the highest ACE.
The numerical results for the external rates at NCC condition are shown in Fig. \ref{fig2}C, exhibiting that the optimal conditions can be realized based on the prediction of the $M$ factor from the fabrication characterization.
With a fairly small $M$, the optimized $\kappa/\gamma$ for two modes are closed to unity, corresponding to the critical coupling condition in the linear system. 
As the nonlinear process strengthens, the required $\kappa/\gamma$ experiences a large increase and exceeds dozens, under which the upper limit of ACE surpasses 50$\%$.

\subsection*{Device}

\begin{figure}
\centering
\includegraphics{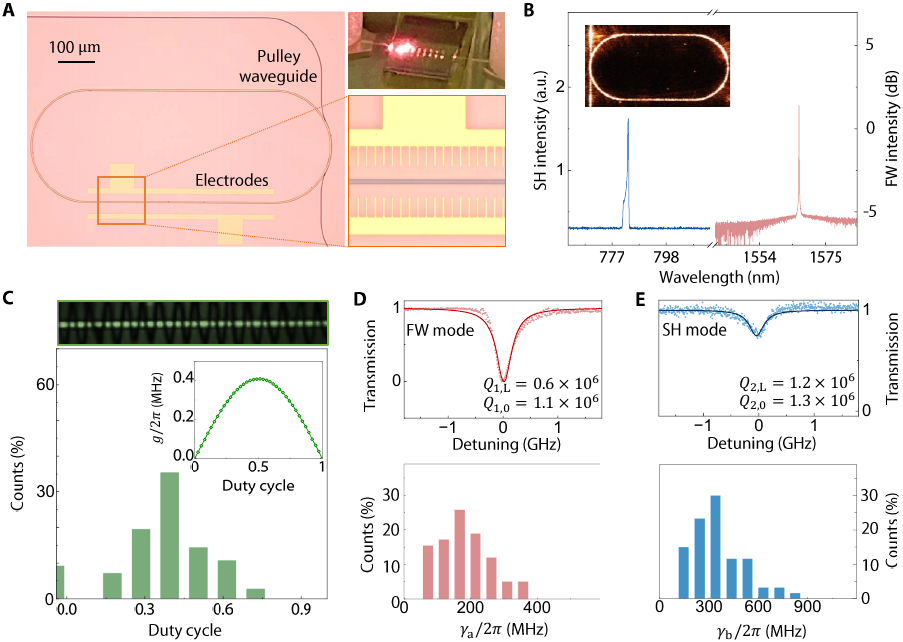}
\caption{\textbf{Fabrication-limited properties of PPLN-integrated microresonators.} 
\textbf{(A)} Microresonators fabricated on an LN chip and its SHG process. 
\textbf{(B)} Optical spectra of the pump and SH lasers. Inset: Microresonator image demonstrating SHG.
\textbf{(C)} Characterization of poling quality. Upper panel: Second-harmonic (SH) microscopy images of the PPLN. Lower panel: Distribution of duty cycles. Inset: Calculated single-photon nonlinear coupling strength $g$ as a function of duty cycle. 
Distribution of intrinsic linewidths for microresonators near 1560 nm FW \textbf{(D)} and 780 nm SH modes \textbf{(E)}. 
Insets: Representative transmission spectra of FW and SH modes under the under-coupled conditions, with the linewidths extracted from Lorentzian fits. }
\label{fig3}
\end{figure}

The devices employed in this study are racetrack microresonators fabricated on X-cut thin-film lithium niobate (Fig.~\ref{fig3}A). The fabrication process is detailed in the Materials and Methods section. The waveguides have a width of 1.8 $\mu\text{m}$, an etching depth of 400 nm, and a sidewall angle of $60^\circ$. The straight waveguide segment of the racetrack is oriented perpendicular to the optical axis, ensuring that the transverse-electric (TE) modes are extraordinarily polarized, thereby leveraging the largest nonlinearity tensor element ($d_{33} = \text{19.5}$ pm/V) \cite{LNtensor}. The FW and SH modes are set to 1560 nm and 780 nm, respectively, both selected as fundamental TE modes to maximize spatial overlap.

To compensate for the azimuthal number mismatch arising from dispersion, periodic poling with a period of $4.3 \mu\text{m}$ is introduced in one arm of the microresonator. 
Optimal quasi-phase matching is achieved when the inverted domains have a duty cycle of 0.5. 
We assess the poling quality using SH microscopy (Fig.~\ref{fig3}C). 
The dark lines between green regions correspond to domain walls between inverted domains. 
By measuring the spacings between these domain walls, we determine the duty cycles, which have a mean of 0.38 and a standard deviation of 0.183. Approximately $9\%$ of the poling attempts fail. 
Based on these statistics and the simulation of the modes overlap, the single-photon nonlinear coupling strength is estimated to be $g/2\pi = \text{0.3} \pm \text{0.12}$ MHz. 
Furthermore, the intrinsic linewidths are derived from transmission spectra acquired using tunable lasers at both FW and SH bands. 
Over 200 modes are measured, and their intrinsic linewidth distributions are summarized in Figs.~\ref{fig3}D and E, exhibiting most probable values of $\gamma_\mathrm{a}/2\pi = 181$ MHz and $\gamma_\mathrm{b}/2\pi = 351$ MHz.

The combination of high poling quality and low intrinsic losses provides a foundation for efficient SHG, and the small variations indicate uniformity in the fabrication process which is important for further design. 
As a preliminary test, pumping the microresonator at 1566 nm results in SHG at 783 nm (Fig.~\ref{fig3}B). 

\subsection*{Nonlinear dispersive and dissipative coupling}

\begin{figure}[!b] \centering \includegraphics{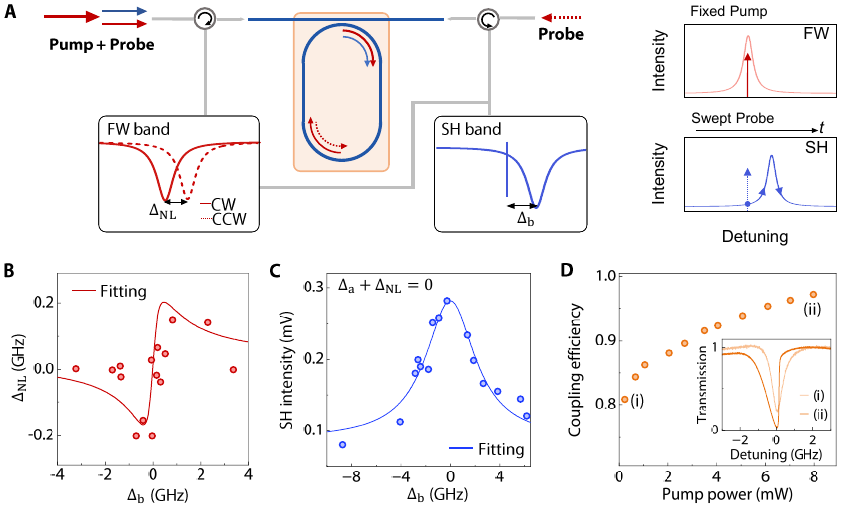} \caption{\textbf{Nonlinear loss and nonlinear frequency shift.} 
(\textbf{A}) Left panel: Experimental setups for measuring the nonlinear frequency shift $\Delta_\mathrm{NL}$ and SH mode detunings $\Delta_\mathrm{b}$. 
Right panel: Schematic of the laser frequency for respectively the FW and SH bands to measure $\Delta_{\rm{b}}$.
(\textbf{B}) Nonlinear frequency shift versus SH mode detuning, measured with the pump resonant to the FW mode ($\Delta_\mathrm{a} + \Delta_\mathrm{NL} = 0$). 
(\textbf{C}) SH power versus SH mode detuning with the pump wavelength aligned to the FW mode. The pump power is set to 1.6 mW.
(\textbf{D}) Coupling efficiency of the FW mode at various pump powers under the over-coupled condition. 
Inset: Transmission spectra of the FW modes at (i) and (ii) power levels.
} \label{fig4}
\end{figure}

The nonlinear frequency shift $\Delta_{\rm{NL}}$ is characterized using a dichromatic pump-probe technique (Fig.~\ref{fig4}A). 
Given that various conventional nonlinear processes shift the mode, we utilize the frequency difference of the bi-directional FW modes to extract the frequency shift induced by the SHG process.
The main part of the mode frequency difference is attributed to SHG-induced dispersive modulation due to the phase-matching condition \cite{tangstrong}, as only the mode propagating in the same direction as the pump laser can be modulated.
Besides, both two modes in the counter-propagating direction carry equal contributions from the global thermo-optic and photorefractive modulations, thus resulting in scarcely frequency difference.
Experimentally, with the pump laser fixed to resonate with the FW mode, we employ weak counterpropagating probe lasers with a scanned frequency in the FW band to obtain the frequency differences.
Besides, an additional probe laser interrogates the SH mode, and by comparing its frequency with the beat note from the SH laser, the detuning $\Delta_\mathrm{b}$ is determined (right panel of Fig. \ref{fig4}A). 
Notably, $\Delta_\mathrm{b}$ can be further adjusted by varying the temperature of the microresonator due to the thermo-optic effect \cite{PPLN250000}.
Therefore,the dependence of $\Delta_\mathrm{NL}$ on $\Delta_\mathrm{b}$ is measured (Fig.~\ref{fig4}B), with the results following the prediction of Eq.~\ref{Eq:nfreqshift}, where the nonlinear frequency shift vanishes with $\Delta_\mathrm{b}$ approaching zero.
There exists a small deviation due to the self- and cross-phase modulations for two counterpropagating lights \cite{2017symmetry,caoboprl}, which contributes slightly.
This frequency modulation manifests as both the Kerr-type and anti-Kerr-type effects, providing an additional control beyond conventional Kerr phase modulation \cite{focusing}.
Meanwhile, we simultaneously monitor the SHG power, which also reaches its maximum value when the SH detuning is zero (Fig. \ref{fig4}C). 
These measurements validate the ideal frequency relations between the pump and the modes for maximal ACE, which are accessible by thermally tuning the microresonator.

To elucidate the nonlinear loss, we experimentally measure the transmission spectra of the FW mode at varying pump power levels. 
In an under-coupled FW design ($\kappa_\mathrm{a} < \gamma_\mathrm{a}$), increasing pump power enhances nonlinear effective intrinsic losses, further increasing the transmitted power at resonance. 
Conversely, in an over-coupled design ($\kappa_\mathrm{a} > \gamma_\mathrm{a}$), higher pump power would drive the FW mode toward critical coupling and result in near-zero transmission at resonance. 
We define coupling efficiency as the fraction of pump power coupled into the microresonator, quantified by the mode depth in the transmission spectra. 
Its dependence on pump power is exhibited in Fig.~\ref{fig4}D, indicating that as the nonlinear process strengthens, the over-coupled design facilitates achieving the NCC condition.

\subsection*{Measurement of ACE}

\begin{figure}[t] 
\centering 
\includegraphics{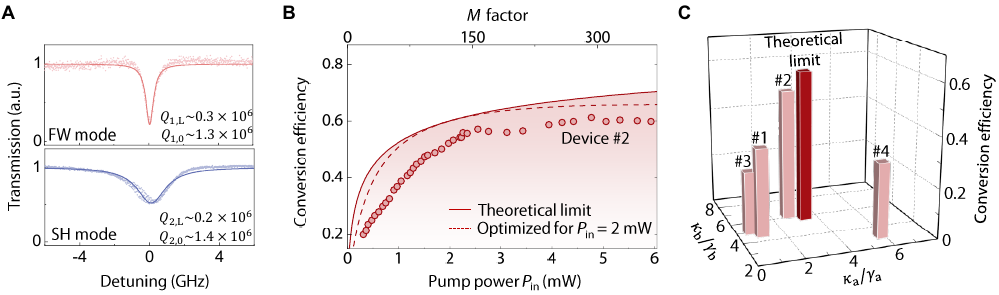} 
\caption{{\bf Coupler design and ACE measurements.}
(\textbf{A}) Transmission spectra of the FW mode and SH mode.
External coupling rates are determined through Lorentzian fitting.
(\textbf{B}) Measured ACE as a function of pump power.
The theoretical limit of ACE under each pump power is plotted as the red solid curve.
Red dashed curve presents the theoretical ACE under the external coupling conditions optimized for a pump power of 2 mW.
(\textbf{C}) Measured ACE for various external coupling rates. 
$M$ is set to 55 across all four devices.
The theoretical limit as well as the required external coupling rates are indicated by dark red rectangular prism.
}
\label{fig5} 
\end{figure}

We design the external coupling rates to achieve optimal ACE, which is given by Eq. \ref{Eq:GCC}.
We employ pulley couplers \cite{pulley,pulleynp1,pulleynp2}, where the top width ($w$), length ($\theta$) of the coupler, and its gap with the microresonator ($d$) can be tuned. 
This provides substantial design flexibility for controlling the external coupling rates at both FW and SH bands. 
For example, we aim to achieve optimal ACE at a pump power of 2 mW, and the $M$ factor is calculated to be 55 based on the averaged device properties. The ideal external coupling rates are $\kappa_\mathrm{a}/2\pi = \text{656}$ MHz and $\kappa_\mathrm{b}/2\pi = \text{2235}$ MHz. Using a pulley waveguide with $w = \text{300}$ nm, $\theta = \text{63}^\circ$, and $d = \text{300}$ nm, we attain external coupling rates close to the ideal settings (Fig.~\ref{fig5}A). 
The measured ACE of this device reaches 52$\%$ at the pump power of 2 mW pump power, approaching the ultimate efficiency for the corresponding $M$ factor (Fig.~\ref{fig5}B).
Increasing the pump power further enhances ACE, reaching 61.3$\%$ at 4.7 mW input power. 
The device is optimized for 2 mW input power, thus, these values fall due to the increasing deviation from the NCC condition at higher pump powers.

The relationship between the measured ACE and the coupling condition toward the NCC condition is summarized by four different devices in Fig. \ref{fig5}C. 
These devices operate at the same predicted $M$ factors but the realized external coupling conditions perform differently.
It shows that the second device of which the coupling conditions are closest to those predicted by Eq.~\ref{Eq:GCC} exhibits the highest ACE, approaching the theoretical limit (60$\%$) under 2 mW pump power. 
Others perform a weaker SHG process due to the larger deviations from the NCC condition.

\subsection*{Discussion} 

Our work addresses the fundamental challenge of overcoming dispersive and dissipative suppression in microcavity-enhanced nonlinear frequency conversion. 
This is achieved through the introduction of the NCC condition, demonstrating that the system can reach a critical state with efficient energy transfer from the FW mode to the SH mode.
Building upon this, we develop a universal protocol for the ACE limit and experimentally achieve a record-high efficiency of 61.3$\%$.
This protocol provides a systematic framework for designing high-performance microresonator-based frequency doublers, comprising three steps: (1) predicting the decisive factor $M$ based on pump power and current nanofabrication capabilities; (2) optimizing resonant enhancement and output coupling via nonlinear input-output analysis to maximize ACE; and (3) fine-tuning system parameters, including laser and mode frequencies, using a dichromatic pump-probe method to mitigate parasitic nonlinear effects such as thermal and photorefractive responses.
This framework is versatile, extending across various nonlinear photonic platforms and applicable to a wide range of processes, including third-harmonic generation, parametric oscillations, and quantum-regime phenomena such as spontaneous parametric down-conversion and spontaneous four-wave mixing.

Notably, this approach enables high-efficiency ACE at low power levels, addressing critical requirements for applications such as self-referencing in microresonator-based optical frequency combs, where the typically microwatt-scale comb lines pose significant challenges.
By leveraging the concept of NCC and achieving unprecedented ACE, our work addresses longstanding efficiency limitations in nonlinear frequency conversion within microcavities. This progress provides a solid foundation for efficient energy transfer in complex optical systems and offers promising opportunities for advancements in both classical and quantum photonic technologies.

\subsection*{Materials and Methods} 

\subsubsection*{Device Fabrication}

We fabricate a periodically poled microcavity on an X-cut LN wafer consisting of a 600\,nm LN layer, a 4.7\,$\mu$m silica layer, and a 0.5\,mm silicon substrate~\cite{fab}. The electrode patterns are defined through electron beam lithography (EBL). Electrodes composed of 20\,nm chromium and 60\,nm gold are deposited via electron-beam evaporation and patterned using a lift-off process. A high-voltage pulse of approximately 600\,V is applied in the $+z$ direction to induce periodic poling.

Waveguide patterning employs EBL with hydrogen silsesquioxane resist as a mask. The defined pattern is transferred to the LN film using argon plasma etching in an inductively coupled plasma reactive ion etching system. Following etching, the chip is immersed in an HF buffer solution to remove residual resistance. To eliminate redeposition, the chip is treated in a solution of H$_2$O, NH$_4$OH, and H$_2$O$_2$ in a 1:2:2 ratio for 20 minutes at 80\,$^\circ$C. Optical-quality facets of the pulley waveguide are prepared by mechanical cleaving to ensure efficient fiber-chip coupling. Fiber-to-chip coupling losses are measured to be 8.67\,dB, and 11.17\,dB per facet for the fundamental and second harmonic bands, respectively.

\subsubsection*{Domain Characterization}

We have carefully designed and fabricated PPLN devices, but the morphology of the inversed domains greatly affects the final conversion efficiency. 
Therefore, accurate characterization of the domain structure is principal for high-efficiency PPLNs. 
To present the quality of the inverted domain without damage, we built a second harmonic microscope imaging system to distinguish the boundaries of the inverted domain region by phase interference of signal light. This system is used for high-resolution and rapid visual demonstration of the domain structure of PPLN, including the position of the domain wall and the duty cycle. The femtosecond light source from the fiber laser  (pulse width$\sim$82 fs, the repetition rate is 34.2 MHz, and central wavelength is 1035 nm) is excited by linear polarization and then irradiated to the incident pupil of the objective lens by the infinite correction 4f system. The FW is focused onto the X-cut lithium niobate sample by the objective lens (Olympus, NA = 0.7), and the back-reflected SH (central wavelength$\sim$517.5 nm) returns along the original path, separated from the fundamental at the dichroic mirror (Thorlabs, DMLP735B), and collected by a photomultiplier tube (Thorlabs, PMT1001). 
We imaged the periodic characteristics of the observed domains under a 20x optical lens. High-intensity areas correspond to high imaging detection efficiency, and vice versa. 
This solution allows us to quickly acquire poling results in situ and without any destruction, which will be conducive to the poling process.

\bibliography{ref}
\bibliographystyle{sciencemag}

\section*{Acknowledgments}
We acknowledge helpful discussions with A. Gao, J.-h. Chen, C.-H. Dong and C.-S. Tian. 
\paragraph*{Funding:}
This project was supported by the National Key R\&D Program of China (Grant No. 2021ZD0301500, 2023YFA1407104), the National Natural Science Foundation of China (Grants No. 12293051, 12322411, 92250302, 12174010), and the High-performance Computing Platform of Peking University. 
\paragraph*{Author contributions:}
Y.-F.X., F. B., and Q.-T.C. conceived the project. Z.-Y.W. implemented the experiments and collected and analyzed the experimental data. X.W. implemented the fabrication and provided the samples. Z.-Y. W. drafted the manuscript, with revisions from  Q.-F.Y., Y.H., X.W., X.X., C.Y., Z.H., F.B., Q.-T.C., and Y.-F.X.
\paragraph*{Competing interests:}
The authors declare that they have no competing interests.
\paragraph*{Data and materials availability:}
All data needed to evaluate the conclusions in the paper are present in the paper or the Supplementary Materials.

\subsection*{Supplementary Materials}
Supplementary Text\\
Figs. S1 to S9\\
Table S1\\
References\\
Movie S1\\

\newpage

\end{document}